\documentstyle[twoside,fleqn,espcrc2,axodraw,psfig]{article}

\def\Journal#1#2#3#4{{#1} {\bf #2}, #3 (#4)}

\def\NPB{{\em Nucl. Phys.} B}
\def\PLB{{\em Phys. Lett.}  B}

\def\PRD{{\em Phys. Rev.} D}  
\def\ZPC{{\em Z. Phys.} C}

\def\ra{\rightarrow}

\def\be{\begin{equation}}
\def\ee{\end{equation}}
\def\bea{\begin{eqnarray}}
\def\eea{\end{eqnarray}}
\def\ba{\begin{eqnarray}}
\def\ea{\end{eqnarray}}

\newcommand{\AmS}{{\protect\the\textfont2
  A\kern-.1667em\lower.5ex\hbox{M}\kern-.125emS}}

\title{The forward photon production and the gluonic content of the real and
 virtual photon at the HERA collider\thanks{Presented at the International Conference
on the Structure and Interactions of the Photon, PHOTON'99, 23-27 May 1999,
Freiburg im Breisgan, Germany.}}

\author{Maria Krawczyk 
\address{Institute of Theoretical Physics, 
	Warsaw University, \\ 
	ul. Ho\.za 69, 00-681 Warsaw,  Poland}%
	\thanks{Supported in part by the Polish Committee for Scientific
Research, Grant No 2P03B18410 (January-June 1999) and 2P03B01414 (1999).}                                    
and Andrzej Zembrzuski
$^{\rm a}$\thanks{Supported in part by the Polish Committee for Scientific
Research, Grant No 2P03B18410 (January-June 1999) and Interdisciplinary
Centre for Mathematical and Computational Modelling, Warsaw University,
Grant No G16-10 (1999).}}                                    
       
\begin{document}
\titlepage
\begin{flushright}
IFT 99/16 \\
{\bf hep-ph/9912368}
\end{flushright}
\vspace*{1.cm}
\centerline{\bf {\huge {{The forward photon production and}}}}
\vskip 0.4cm 
\centerline{\bf {\huge {the gluonic content
of the real and virtual photon}}}
\vskip 0.4cm 
\centerline{\bf {\huge {at the HERA collider}}}
\vskip 2.cm
\centerline{\Large Maria Krawczyk$^{\dagger}$ 
and Andrzej Zembrzuski$^{\ddagger}$}
\centerline{Institute of Theoretical Physics, Warsaw University, Poland}

\vskip 2cm
\centerline{\bf {\Large {Abstract}}}
\vskip 0.5cm 
The discussion on the production of  prompt photons with $p_T$ of a few GeV
in the tagged process $ep\rightarrow e \gamma X$
with small $Q^2$ (DIC process) at the HERA collider is presented. 
Photons produced in the forward (proton) direction 
are mainly originating from subprocesses involving interactions of the
gluonic content  of the exchanged photon. The large enhancement over 
the Born term (direct photon) is found up to a factor of 35 for a real
photon and up to 5 for a virtual photon with a squared virtuality 
1~GeV$^2$. It gives a possibility of 
extracting  a  gluonic density of the real and of the virtual photon.
The  BFKL approach to the description of the forward particle
production is shortly discussed.

\vskip 1.6cm
~\newline
Presented at the International Conference
on the Structure and Interactions of the Photon, PHOTON'99, 23-27 May 1999,
Freiburg im Breisgan, Germany.

\vskip 1cm
~\newline
{\footnotesize $^{\dagger}$Supported in part by the Polish Committee for 
Scientific
Research, Grant No 2P03B18410 (January-June 1999) and 2P03B01414 (1999).
\newline
$^{\ddagger}$Supported in part by the Polish Committee for Scientific
Research, Grant No 2P03B18410 (January-June 1999) and Interdisciplinary
Centre for Mathematical and Computational Modelling, Warsaw University,
Grant No G16-10 (1999).}


\begin{abstract}
The discussion on the production of  prompt photons with $p_T$ of a few GeV
in the tagged process $ep\rightarrow e \gamma X$
with small $Q^2$ (DIC process) at the HERA collider is presented. 
Photons produced in the forward (proton) direction 
are mainly originating from subprocesses involving interactions of the
gluonic content  of the exchanged photon. The large enhancement over 
the Born term (direct photon) is found up to a factor of 35 for a real
photon and up to 5 for a virtual photon with a squared virtuality 
1~GeV$^2$. It gives a possibility of 
extracting  a  gluonic density of the real and of the virtual photon.
The  BFKL approach to the description of the forward particle
production is shortly discussed.
\vspace{1pc}
\end{abstract}

\maketitle

\section{Introduction} 
\subsection{The DIC process at HERA}

The production of the prompt photon with a large transverse momentum   
in $ep$ collision
under antitagging or untagging conditions
is dominated by events with almost real photons mediating 
the $ep$ interaction, $Q^2\approx 0$. So in practice the 
photoproduction of the prompt photon, the Deep Inelastic Compton 
(DIC) scattering, is considered. The observed final $\gamma$
arise directly from a hard subprocess
or from a fragmentation process,
where  $q$ or  $g$ 'decays' into $\gamma$. 

This DIC process with two real photons first
offered an opportunity to test  the Parton Model 
idea \cite{DIC-PM} and later the Quantum Chromodynamics with LO 
and NLO accuracy \cite{font,duke,aure,mk,bks,prob,mk4,aurenche,gs}.
Next, the isolation restriction of the final photon needed in
experimental analyses \cite{h1,z1,z2} was imposed in the NLO
calculations \cite{gordon4}.
Recently the NLO approaches 
with the isolation of the final photon and other 
experimental cuts \cite{gordon7,my,my2,phd}
were applied successfully to describe the data measured at the HERA collider
\cite{z1,z2}.

The sensitivity of the DIC process to test the parton densities in the photon 
(and in the proton) was  realized and studied for the HERA collider already 
ten years ago \cite{mk,bks,prob,mk4,aurenche}. 
The conclusion 
was reached that the processes initiated by the gluons
from the photon dominate in events where the final photon 
is produced in the forward direction, i.e. in the 
direction of the initial proton.
This is a specific kind of an inverse Compton process
characterized by very energetic final photons \cite{mk,bks,prob,mk4}. 
The tagging condition, limiting the virtuality 
and the energy of the exchanged photon, allows to separate the individual 
contributions, especially these with the gluonic content of the photon
\cite{prob,mk4} (see also \cite{gs}).

The DIC process with so called resolved photon (i.e. interacting
via its partons) takes place also for the {\underline {virtual photon}}.
This occurs when the (positive) virtuality of the exchanged photon, 
denoted below by $P^2$ $^($\footnote{It is convenient to denote the
virtuality of the photon-target by $P^2$, while use $Q^2$ if the
photon plays a role of a probe.}$^)$, differs from zero and yet 
is still small compared to the hard scale of the partonic 
subprocesses given  by the 
transverse momentum of the final photon, $P^2\ll p_T^2$. Then
an opportunity arises to test partonic (gluonic) structure 
of the virtual photon in a similar manner as 
for a (almost) real photon
mentioned above
\cite{azem,probv}. 

The recent HERA data for the jet production processes 
allowed so far to extract a combination of parton densities
- the so called effective parton distribution
in the virtual photon \cite{eff-virt-data}. 
We argue that the photon production (DIC process)
may play an complementary role in testing the structure of virtual photon
\cite{survey}. Although it appears at lower rate the 
production of photon 
has an advantage being dominated in the forward direction
 by only one, a gluon initiated subprocess.
Therefore the DIC process may allow to separate and 
extract the fundamental, although less known experimentally,
gluonic  density in the virtual photon. 

\subsection{Forward particle production as a test of QCD dynamics 
at small $x_{Bj}$}
Recently a forward jet and
particle production, among them $\pi^0$ and $\pi^{\pm}$,
has been studied at HERA \cite{H1-for,ZEUS-for} in  order to test
QCD dynamics at small $x_{Bj}$. The question is
what is the proper type of partonic evolution (BFKL, DGLAP or CCFM?)
in the chain between the virtual photon and the proton
for the DIS events at small $x_{Bj}$.
In such analysis the positive virtuality of the exchanged photon,
$P^2$, plays  a role of the standard DIS $Q^2$ variable.
It is well known that the BFKL evolution corresponds to 
not strongly ordered in $p_T$
parton emission and leads to larger transverse momenta of final gluons
 than the DGLAP evolution
based on  the collinear gluon emissions.
Therefore  in  search for a signal of the BFKL evolution  
the transverse momentum of the final forward particle
is usually  chosen to be large and of order of $ P^2$.

The resolved virtual photon contribution
leads to  BFKL-type final state configurations,
as it was pointed out in \cite{jung}.
Although  such contributions
is suppressed 
for $p_T^2\sim P^2$  (lack of corresponding large logarithms 
present in the standard DGLAP description of the virtual photon 
structure, where $p_T^2\gg P^2$), one should be aware that
the forward particle  production may arise from a
 mechanism not obviously included in the BFKL
evolution equation \cite{jung,kram-pott}.

The newest ZEUS data \cite{ZEUS-for} for the forward jet
 production studied in the range $10^{-2}< {{p_T^2}/ {P^2}} < 10^2$
show that only model which includes  resolved photon components, with
a hard scale $\tilde Q^2=P^2+p_T^2$, describe the data in the whole kinematic
range. On the other hand a new analysis \cite{jung-99} shows that CCFM 
approach can describe properly the forward jet production at HERA as well,
while the H1 data for forward  $\pi^0$ \cite{H1-for} seems to support
 a modified  (i.e. with a limited $p_T$ phase space) BFKL   
evolution equation
\cite{kwiec-pi}. 

~\newline\newline
In this presentation we focus on
the production of very energetic forward photons with $p_T \sim$ 5 GeV
in the DIC process for $P^2\ll p_T^2$ and we show that it may offer an
opportunity to probe the gluonic content of the virtual photon at HERA.
\footnote{The analysis of the forward $\gamma$ production at HERA   
as a signature of the BFKL evolution can be found in \cite{kwiec}.}
 
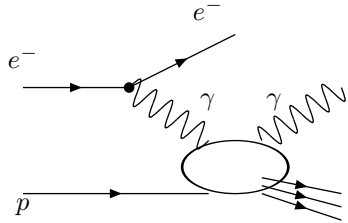
\begin{figure}
\begin{center}
\begin{picture}(120,120)(0,20)
\ArrowLine(0,90)(40,90)
\Text(0,102)[]{\mbox{$e^-$}}
\Vertex(40,90){2}
\ArrowLine(40,90)(80,110)
\Text(70,120)[]{\mbox{$e^-$}}
\Photon(40,90)(70,70){5}{5}
\ArrowLine(0,50)(70,50)
\Text(0,45)[]{\mbox{$p$}}
\Text(70,85)[c]{\mbox{$\gamma$}}
\Text(95,85)[c]{\mbox{$\gamma$}}
\ArrowLine(90,56)(120,50)
\ArrowLine(90,53)(120,45)
\ArrowLine(90,50)(120,40)
\Photon(90,70)(120,90){5}{5}
\Oval(80,60)(10,20)(0)
\end{picture} \end{center}
\vspace{-1.5cm}
\caption{The DIC process}
\label{dic}
\end{figure}

\section{Probing parton densities in $\gamma^*$ in the DIC process
at HERA}


We continue our study \cite{probv} of the inclusive DIC  process
\be
ep \ra e \gamma X
\ee
for a relatively small momentum transfer between the initial
and the final electron, 
where the 
exchange of a virtual photon dominates over the $Z$ boson exchange.
We will limit ourselves to events with the tagged electrons,
so the energy and 
(positive) virtuality of the exchanged photon $-p^2=P^2$ are known.

The first attempt to describe the DIC process at HERA using the 
structure of the virtual photon can be found in \cite{azem},
where the Equivalent Photon Approximation (EPA) 
approach was compared with the calculation
involving direct and resolved virtual photons.
Next we have examined the usefulness of DIC process  
to study at the  HERA collider the structure of a 
virtual photon,
in particular its gluonic content \cite{probv}.
Here we extend this analysis by comparing 
compare different contributions to the DIC cross section 
in the $ep$ LAB reference frame
and by studying  a wider virtuality range: $ P^2 $ from 0 to 2.5 GeV$^2$. 

\subsection{Born term and subprocesses involving resolved $\gamma^*$}

We consider  
the virtual photon - proton interaction leading to the 
production of the large $p_T$ photon. 
It corresponds to the following direct subprocess 
at the lowest order (Born level):
\be
\gamma^* q_{p}\ra\gamma q,
\ee
with the initial and final photon  interacting  directly 
with a quark from the proton. The Born process
 dominates at $very$ $large$ $p_T$ $\sim 
\sqrt{S_{ep}}/2$ $(x_T = {2p_T\over \sqrt{S_{ep}}} \sim 1)$,
while in the $moderate$ $p_T$ region which we will consider here, 
i.e. for $\Lambda_{QCD}^2\ll P^2 \ll p_T^2\ll{S_{ep}}/4$,
resolved virtual photon processes become important.
There are three   types of  LO subprocesses 
involving the partonic constituents  of the {\underline {initial}} 
and/or {\underline {final}}
photons in DIC process:
\begin{itemize}
\item
with single resolved initial photon
\be
g_{\gamma^*} q_p\ra\gamma q,
\ee
\be
q_{\gamma^*}g_p\ra\gamma q,
\ee
\be
q_{\gamma^*}\overline q_p\ra\gamma g
\ee
\be
\overline q_{\gamma^*} q_p\ra\gamma g
\ee
\item 
with single resolved final photon
(fragmentation into the photon)
\be
\gamma^* g_p\ra q {\bar q} 
\ee
\be
\gamma^* q_p\ra g { q} 
\ee
\item with double resolved photons
\be
g_{\gamma^*} g_p\ra g g
\ee
\be
q_{\gamma^*} g_p\ra q g, etc.
\ee
\end{itemize}
In this talk we limit ourselves to the processes
involving single resolved initial photon. We expect to see 
the effect due to the gluonic content in the 
virtual photon by looking into a final photon produced in the 
forward direction. The remaining subprocesses
are known \cite{probv} to give smaller contributions in this
region.  
\subsection{Calculation of the cross section}
The differential cross section for deep inelastic 
electron-proton scattering
with a photon in the final state, eq. (1), can be written in 
the following form:
\bea
E_eE_{\gamma}{d\sigma^{ep\ra e\gamma X}\over 
d^3p_ed^3p_{\gamma}} = 
\Gamma_T\Bigl( E_{\gamma}{d\sigma^{\gamma^*p\ra 
\gamma X}\over 
d^3p_{\gamma}}\Bigl)|_T+ \nonumber\\
+\Gamma_L \Bigl(
E_{\gamma}{d\sigma^{\gamma^*p\ra \gamma X}\over
d^3p_{\gamma}}\Bigl)|_L,~~~~
\eea
where $E_e(E_{\gamma})$ and $p_e(p_{\gamma})$ are the energy 
and the momentum of the final state electron (photon).
Coefficients $\Gamma_T$ and $\Gamma_L$ 
(functions of the energy and momentum of the electron in 
the initial and final state) can
be interpreted as the probability of emitting by the initial electron a
virtual photon polarized transversely and longitudinally. 

Since the cross section for the 
reaction $ep\ra e\gamma X$ is dominated by 
the exchange of the photon with small virtuality, 
one  can neglect a contribution due to  
the longitudinal polarization. (See also \cite{ula}, where the role of the
longitudinal photons is discussed for the Born term, 
and where it was found that the longitudinal
polarizations can be neglected
even for $P^2$ larger than studied here).
Assuming that the exchanged photon has only the
transverse polarization we obtain:
\ba
E_eE_{\gamma}{{d\sigma^{ep\ra e\gamma X}}
\over{d^3p_ed^3p_{\gamma }}}
=\Gamma_T E_{\gamma}{{d\sigma^{\gamma^*p\ra \gamma X}}\over 
{d^3p_{\gamma}}}|_T.
\ea
where we include contributions from subprocesses (2)-(6).
In the calculations of the cross section for the $\gamma^* p$ 
collision functions describing parton densities
appear. They depend
on  a hard scale $\tilde Q^2\sim p_T^2$, a  parton virtuality $P^2$
and $x_{\gamma}$ - the part of the photon momentum taken by the  parton.

It is worth to emphasize that while calculating
the cross section for individual 
subprocesses we take into consideration the
virtuality ($P^2$) of the photon emitted by 
the electron as it follows from the kinematics of the process. 

Note that the minimum value of the $x_{\gamma}$ is reached 
when the photon is produced in the forward (proton) direction.
This fact influences the shape of the cross section as a function of
rapidity what is discussed below.

\subsection{Results}
The resolved initial photon contributions to the DIC process are
calculated and compare  to the Born (direct) term for the HERA collider.
The squared energy of the HERA accelerator ($S_{ep}$) is taken as
 98400 GeV$^2$ and the cross section is calculated 
for the transverse momentum 
of the final photon equal to 5 GeV and for the fixed energy 
of the exchanged photon:  $E_{\gamma^*}=y E_e$, with $y = 0.5$.

In our calculations
the GRS (LO) parton parametrizations  for the parton  
distributions in the  virtual photon \cite{grsparam} and
the GRV (LO) set of the quark and the gluon densities
for the proton \cite{grv} were used. 
The number of flavours is assumed $N_f = 3$ -
 the maximum number of active massless quarks in the GRS
parametrization. We take the QCD parameter $\Lambda_{QCD}=0.2$ GeV
and the hard scale equal to the transverse momentum of the final photon
$\tilde Q=p_T$.

The following cross section was studied:
\be
E_{\gamma}{d\sigma^{ep\ra e\gamma X}\over  
{d^3p_{\gamma} dP^2 dy}}=\\
{{1}\over{2 \pi}}{{E_{\gamma}}\over{p_T^2}} {d\sigma^{ep\ra e\gamma X}\over
{dE_{\gamma}dY dP^2 dy}},
\label{label}
\ee
where $Y$ denotes the $ep$ LAB rapidity of the produced photon
(its positive value corresponds to the proton direction).

The cross section dependence on rapidity was tested 
using various values of $P^2$ between $\sim 0$ and 2.5 GeV$^2$. 
The results are presented in Figures 2 and 3.
Fig.~2 shows the ratio of the cross section (\ref{label}) for the 
subprocess $g_{\gamma^*} q_p\ra \gamma q$ to the Born contribution
as a function of the rapidity $Y$. 
For comparison also the ratios of 
$q_{\gamma^*}g_{p}\ra \gamma q$ and $q{\bar q}\rightarrow\gamma g$
subprocesses to the Born term are plotted.

The virtuality $P^2$ for the reaction $g_{\gamma^*} q_p\ra \gamma q$ 
varies from $10^{-7}$ GeV$^2$ 
$^($\footnote{Due to smooth behaviour of the GRS parametrization  
in the limit $P^2\ra 0$ we were able
to  perform the calculation also below $\Lambda^2_{QCD}$.}$^)$
to 1 GeV$^2$,
while for the other two subprocesses involving the quark or antiquark
content of $\gamma^*$ only the case of $P^2$=0.1 GeV$^2$ is presented
(results for higher $P^2$ are even smaller).

The clear dominance of the process $g_{\gamma^*} q_p\ra \gamma q$
over the Born contribution and over the other two types of subprocesses 
with a single resolved virtual photon
is seen in the forward direction. 
Even for $P^2$=1 GeV$^2$ the corresponding cross
section ratio reaches maximum value of about 6 at $Y_0$=4.8.
The position of the peak is almost independent of $P^2$ and it 
is related to
the smallest value of $x_{\gamma}$ tested in the subprocess
($min$ $x_{\gamma}\sim x_T^2$ and $Y_0^{CM}\sim -\ln x_T$, 
see \cite{mk}).

The same cross section ratio, now only for the
dominating subprocess $g_{\gamma^*} q_p\ra \gamma q$,
as a function of the virtuality $P^2$ for three different values 
of positive  $Y$ 
is presented in Fig.~3.
The large enhancement of the contribution due to the
gluonic content of the virtual photon over the Born term
is seen in the proton direction for the 
considered virtuality range. 

\begin{figure*}[htb]
\vspace{9pt}
\psfig{file=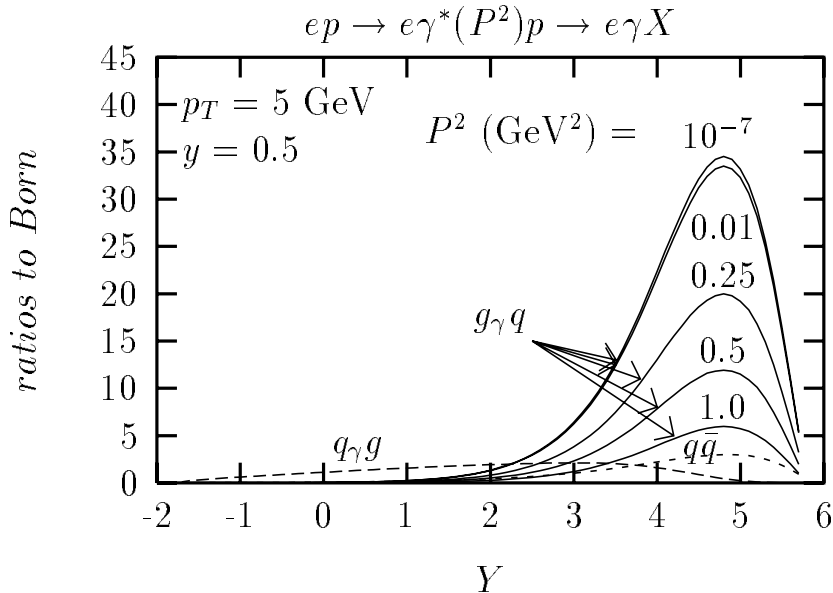,height=60mm}
\caption{Ratios of the cross sections (see eq. 13) 
for the reaction $ep \ra e\gamma X$, with $y$=0.5 and $p_T = 5$ GeV.
The ratios of the contributions due to 
$g_{\gamma^*}q_p\ra \gamma q$ (solid lines),
$q_{\gamma^*}g_p\ra \gamma q$ (long-dashed line)
and ${\bar q_{\gamma^*}} q_p$, 
$q_{\gamma^*} {\bar q_p}\ra \gamma g$ (short-dashed line)
to the Born cross section are presented for various
values of the  virtuality $P^2$ as a function of the rapidity $Y$. 
For $g_{\gamma^*}q_p\ra \gamma q$ the results correspond to $P^2$=
$10^{-7}, 0.01,0.25,0.5 ~{\rm and}$ 1 GeV$^2$. 
For $q_{\gamma^*}g_p\ra \gamma q$ and 
$q_{\gamma^*} {\bar q_p}$, ${\bar q_{\gamma^*}} { q_p} \ra \gamma g$
we take $P^2$= 0.1 GeV$^2$.}
\label{fig:rapt}
\end{figure*}
\begin{figure*}[htb]
\vspace{9pt}
\psfig{file=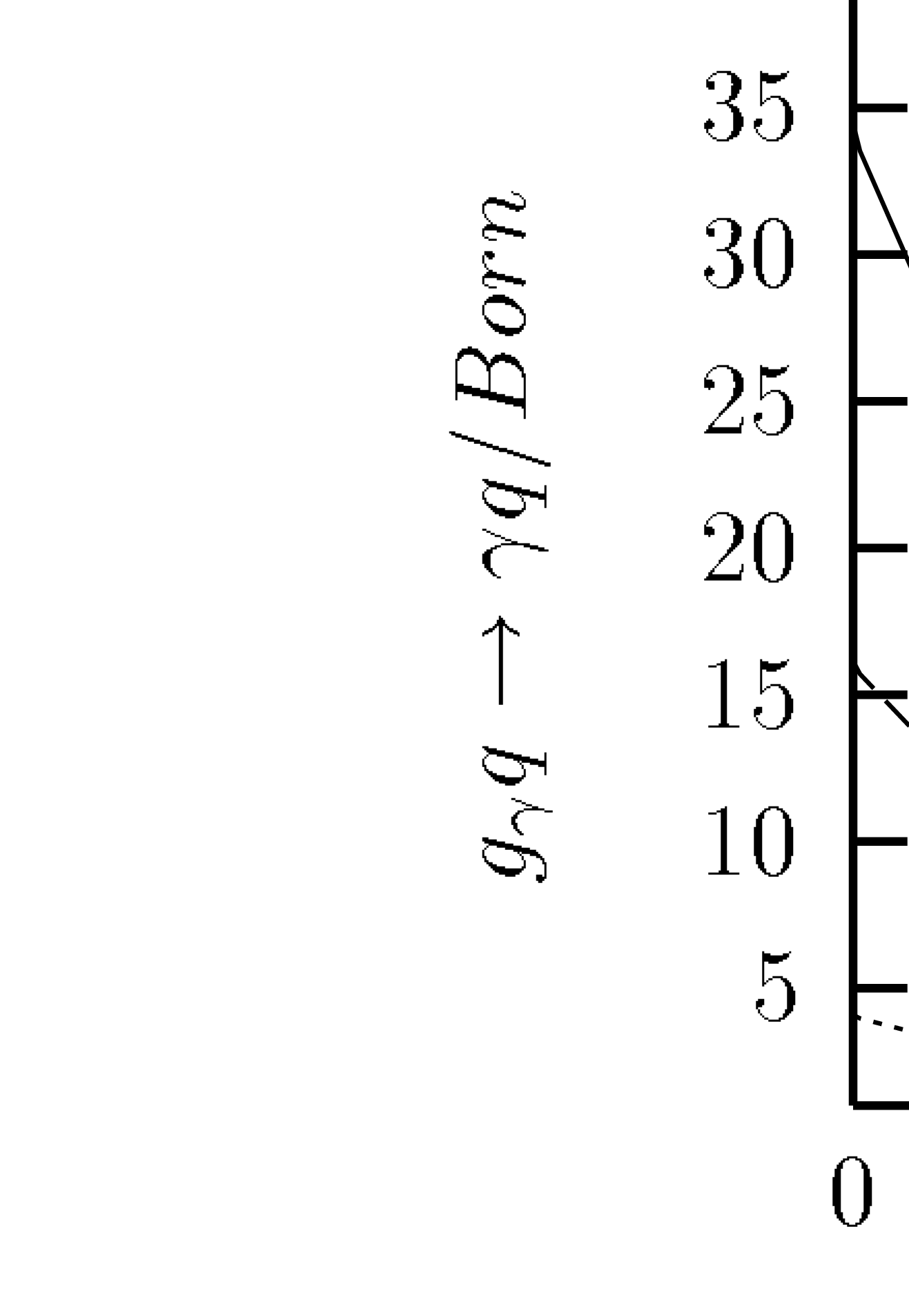,height=55mm}
\caption{ The ratio of the cross sections
${d\sigma^{ep\ra e\gamma X}\over{dE_{\gamma}dY dP^2 dy}}$
for $p_T = 5$ GeV and $Y = 0.5$:
$g_{\gamma^*} q\rightarrow \gamma q$ contribution divided by the Born
contribution. The lines show the results as a function of the initial 
photon virtuality $P^2$. The final photon rapidity $Y$ 
is equal to
2.8 (dotted line), 3.7 (dashed line) and 4.8 (solid line). 
}
\label{fig:virt}
\end{figure*}

This dominance of the $g_{\gamma^*}q_p$ process
is so large that the change of parameters (among them
the number of active flavours) and/or parton 
parametrizations  will not change the main conclusion. 
The conclusion will remain unchanged also
when imposing an experimental isolation restrictions
for the final photon, because the isolation does not 
change the cross section for the dominant subprocesses (2-6)
considered herein\cite{my,my2,phd}.

Note also that    
the interference with the Bethe-Heitler process, discussed in 
\cite{brodsky}, is small 
for $p_T$=5 GeV and
in the region of the rapidity where the 
gluonic content of the virtual photon plays 
a dominant role \cite{pj}.

\section{Conclusions}
We have considered the LO cross section for
tagged DIC events corresponding to the $\gamma^*p$ scattering 
including the direct and resolved
virtual photon subprocesses at the HERA collider.
The GRS$(\gamma^*)$ and GRV$(p)$ parton
parametrizations were used for $N_f$=3. 
The final photon rapidity dependence was studied
for typical values of the variables, $y$=0.5 and $p_T$=5 GeV,
and virtualities of the exchanged photon from 0 to 2.5 GeV$^2$. 
The ratios of single resolved photon 
contributions to the Born contribution were studied as a function of 
the rapidity and as a function of $P^2$.
 
The  contribution due to gluonic content of the virtual photon
was found to dominate over other subprocesses in the direction of 
the initial proton. 
This can have important consequences for the possibility 
of measuring the gluon distribution in 
the virtual photon at HERA. The DIC process
may be treated as complementary to the jet production which 
have allowed recently to extract 
the effective parton density in the virtual photon at HERA. 

Finally, it is worth to mention that the 
forward photon production in DIC process
leads to the configurations typical for 
the study of the BFKL dynamics at small $x_{Bj}$. 

\section*{Acknowledgments}
One of us (MK) would like to thank the organizers of this very fruitful 
conference for their help before, during and after the
conference. A.Z. is grateful to Joanna Trylska for helpful discussions.


\end{document}